\documentclass[12pt,preprint]{aastex}
\usepackage{graphicx}
\pdfoutput=1

\begin{document}

\title{The telegraph approximation for focused cosmic-ray transport 
  in the presence of boundaries}

\author{Yuri E. Litvinenko, Frederic Effenberger\altaffilmark{1}} 
\affil{Department of Mathematics, University of Waikato, P. B. 3105, Hamilton, New Zealand}

\and

\author{Reinhard Schlickeiser} 
\affil{Institut f\"ur Theoretische Physik, Lehrstuhl IV: 
  Weltraum- und Astrophysik, Ruhr-Universit\"at Bochum, D-44780 Bochum, Germany}

\altaffiltext{1}{Present address: Department of Physics and KIPAC, 
  Stanford University, Stanford 94305, USA}

\begin{abstract}
Diffusive cosmic-ray transport in nonuniform large-scale magnetic
fields in the presence of boundaries is considered. 
Reflecting and absorbing boundary conditions are derived 
for a modified telegraph equation with a convective term. 
Analytical and numerical solutions of illustrative boundary problems are presented. 
The applicability and accuracy of the telegraph approximation for focused cosmic-ray
transport in the presence of boundaries are discussed, and 
potential applications to modeling cosmic-ray transport are noted. 
\end{abstract} 


\section{Introduction}

When energetic cosmic-ray particles propagate in the cosmos, 
they often interact with turbulent magnetic fields. The evolution of 
a particle distribution is governed by the Fokker--Planck equation 
(e.g., Schlickeiser 2011, and references therein). 
When pitch-angle scattering is strong enough to ensure that the scale of 
density variation is significantly greater than the particle mean free path, 
the evolution can be approximated as a simpler diffusive process 
(Hasselmann \& Wibberenz 1970; Schlickeiser \& Shalchi 2008). 

The diffusion approximation has been employed for studying 
the acceleration and propagation of energetic particles in 
various astrophysical situations---from the atmosphere of the Sun
and interplanetary space 
(e.g., Bieber et al. 1987; Le Roux \& Webb 2009; Artmann et al. 2011) 
to the interstellar medium 
(e.g., Schlickeiser, 2009; Litvinenko \& Schlickeiser 2011; 
Schlickeiser et al. 2011). Observations reveal large-scale spatial variations 
of the magnetic field in all those locations 
(e.g., Sofue et al. 1986; Sandroos \& Vainio 2007; Dr{\"o}ge et al. 2010). 
The diffusion model has been developed to incorporate the coherent 
particle streaming due to the effect of adiabatic focusing 
in a nonuniform background magnetic field 
(Earl 1981; Beeck \& Wibberenz 1986; see also Litvinenko 2012a, 2012b; 
He \& Schlickeiser 2014, and references therein). 

Propagation of solar energetic particles in interplanetary magnetic fields 
remains a subject of intense research activity 
(e.g., Zhang et al. 2009; Dr{\"o}ge et al. 2010; Qin et al. 2013; 
Laitinen et al. 2013; Wang \& Qin 2015). The theoretical studies are 
motivated by the new data from multi-point spacecraft observations, 
which allow new insights into the physics of the particle
transport (e.g., Dresing et al. 2014; Dr{\"o}ge et al. 2014;
Lario et al. 2014).  Since theoretical modeling is
usually based on numerical solutions, simple analytical approximations 
can complement and guide the simulations.

A general shortcoming of the diffusion approximation is 
that the diffusion equation implies an infinite speed 
of signal propagation, whereas particle speeds are finite, of course. 
A more accurate description may be provided by the telegraph equation 
(Goldstein 1951). This is an equation of hyperbolic type, 
and its solution at long times asymptotically approaches 
the solution of the diffusion equation (Davies 1954). 
The telegraph equation for Brownian motion 
(Brinkman 1956; Sack 1956) 
had been shown to be substantially more accurate 
than the diffusion equation (Hemmer 1961). 
The derivation of a generalized telegraph equation and its applications 
for modeling cosmic-ray transport had been repeatedly considered 
both in the limit of a uniform background magnetic field 
(e.g., Fisk \& Axford 1969; Earl 1974, 1992; 
Gombosi et al. 1993; Schwadron \& Gombosi 1994) 
and in the more realistic case of a spatially varying field 
(e.g., Earl 1976; Pauls \& Burger 1994; 
Litvinenko \& Noble 2013; Litvinenko \& Schlickeiser 2013; Malkov \& Sagdeev 2015). 

Boundaries typically need to be considered in cosmic-ray transport problems 
in both interplanetary and interstellar plasmas 
(e.g., Schlickeiser 2009; Artmann et al. 2011). 
The boundary conditions for both the diffusion equation and 
the standard telegraph equation are well known 
(Masoliver et al. 1992, 1993). 
We are not aware of a published derivation of 
the corresponding boundary conditions for focused particle transport, 
described by a modified telegraph equation. 
In this paper, we develop the telegraph approximation 
for focused cosmic-ray transport in the presence of boundaries. 
The new analytical and numerical results complement those for cosmic-ray 
transport in the absence of boundaries, which we presented in our 
previous study of the telegraph approximation (Effenberger \& Litvinenko 2014). 

Following a brief description of the model (Section 2), 
we derive the reflecting and absorbing boundary conditions (Section 3), 
illustrate the use of the Laplace transform and Fourier series for obtaining 
analytical solutions (Section 4), and compare them with numerical solutions 
of the Fokker--Planck equation (Section 5).

\section{The telegraph approximation for focused cosmic-ray transport}

Spatial non-uniformity of the mean magnetic field leads to 
the adiabatic focusing that results in coherent streaming 
of cosmic-ray particles along the mean field 
(Roelof 1969; Kunstmann 1979).
Earl (1976) derived a modified telegraph equation for 
the focused particle transport in a spatially varying
magnetic field. The equation, however, described 
the coefficient of an eigenfunction expansion rather than the particle 
density that is the physical quantity of interest. 
The problem was recently reexamined, and 
the telegraph approximation was derived for the particle density 
in a spatially varying magnetic field in a weak focusing limit 
(Litvinenko \& Schlickeiser 2013). 
The telegraph approximation has also been obtained 
in a complementary case of an arbitrary constant focusing strength 
and isotropic pitch-angle scattering (Litvinenko \& Noble 2013), 
using an exact expression for the variance of a particle distribution, 
obtained by Shalchi (2011).

A possible physical context for the telegraph approximation is provided 
by the Fokker--Planck equation for focused transport, which describes 
the evolution of the distribution function of energetic particles: 
\begin{equation}
  \frac{\partial f_0}{\partial t}
  + \mu v \frac{\partial f_0}{\partial z}
  + \frac{v}{2L}(1-\mu^2)\frac{\partial f_0}{\partial \mu}
  = \frac{\partial}{\partial \mu} \left( D_{\mu\mu}\frac{\partial f_0}{\partial \mu}\right) 
\label{eq:FPE}
\end{equation}
(Roelof 1969; Earl 1981). 
Here $f_0(z,\mu,v,t)$ is the distribution function 
(gyrotropic phase-space density), $t$ is time, $\mu$ is the
cosine of the particle pitch angle, $v$ is the particle
speed, $z$ is the distance along the mean magnetic field $B_0$,
$L=-B_0/(\partial B_0/\partial z)$ is the adiabatic focusing length,
and $D_{\mu\mu}$ is the Fokker--Planck coefficient for pitch-angle scattering. 

To illustrate the application of the approximation to a model transport problem, 
we consider the simplest physically plausible model. 
Specifically, we assume a constant focusing length $L$, 
and we neglect momentum diffusion, adiabatic cooling, and advection 
with a background flow (say, the solar wind). 
The simplifying assumptions are discussed, for example, by 
Artmann et al. (2011) and Effenberger \& Litvinenko (2014) in the context of 
solar energetic particle transport in interplanetary space.

The distribution function can be expressed as the sum of 
the isotropic density $F_0$ and an anisotropic component $g_0$: 
\begin{equation}
  f_0 (z, \mu, t) = F_0 (z, t) + g_0 (z, \mu, t) , 
\end{equation} 
where
\begin{equation}
  F_0 (z,t) = \frac{1}{2} \int_{-1}^{1} f_0 d\mu . 
\end{equation}
Assuming that $g_0 \ll F_0$, an approximate expression for $g_0$ can be found 
and substituted into Equation (\ref{eq:FPE}), integrated with respect to $\mu$. 
Depending on the accuracy of the expression for $g_0$, the result is either the usual 
diffusion approximation or the telegraph approximation for focused transport 
(e.g., Earl 1976; Litvinenko \& Schlickeiser 2013). 
Here we investigate the modified telegraph equation for the isotropic density: 
\begin{equation}
  \frac{\partial F_0}{\partial t} + 
  \tau \frac{\partial^2 F_0}{\partial t^2} = 
  \kappa_{\|} \frac{\partial^2 F_0}{\partial z^2} + 
  \xi \kappa_{\|} \frac{\partial F_0}{\partial z} , 
\label{eq:telF0}
\end{equation}
where $\xi \kappa_{\parallel}$ is the coherent speed (e.g., Earl 1981), 
$\kappa_{\parallel}$ is the parallel diffusion coefficient, 
$\xi=\lambda_0/L$ is the focusing strength, and $\lambda_0$ is 
the scattering mean free path in the absence of focusing ($\xi=0$). 
Equation (\ref{eq:telF0}) is written in dimensionless form 
by measuring distances in units of $\lambda_0$, 
speed in units of the constant particle speed $v$, 
and time in units of $\lambda_0/v$. 
Although we formally recover the diffusion approximation by setting $\tau=0$, 
in practice $\tau$ is of order unity  in a physically relevant 
parameter range (Litvinenko \& Schlickeiser 2013). 

The coefficients $\kappa_{\|}$ and $\tau$ generally depend on 
the focusing strength $\xi$. In the weak focusing limit $\xi^2 \ll 1$, 
the transport coefficients are given by Equations (9) and (14) in 
Litvinenko \& Schlickeiser (2013) for an arbitrary pitch-angle 
scattering coefficient $D_{\mu\mu}$. For isotropic scattering, 
the telegraph equation has been derived for an arbitrary 
focusing strength $\xi$, and $\kappa_{\|}$ and $\tau$ are given by 
Equations (9) and (24) in Litvinenko \& Noble (2013). The complementary 
expressions agree in the limit of weak focusing and isotropic scattering. 

It is worth noting that the derivation of the telegraph equation 
by Earl (1973) and Litvinenko \& Schlickeiser (2013) involves truncating 
an infinite system, equivalent to the original Fokker--Planck equation. 
A well-known weakness of that approach (Gombosi et al. 1993; Schwadron \& Gombosi 1994) 
has been recently reiterated by Malkov \& Sagdeev (2015). 
Truncating the system of harmonics, derived from the Fokker--Planck equation, 
leads to an error in the coefficient $\tau$---and more generally, 
to an equation which is not correctly ordered. 
The correct way of deriving the telegraph equation relies on 
an asymptotic expansion that yields the diffusion equation in the lowest order 
and the telegraph equation in the next order. 
Gombosi et al. (1993) described the method in detail for the case of 
a uniform background magnetic field and isotropic scattering. 
Of course the correct mathematical procedure is generally preferable. 
In practice, however, the improvement in accuracy may be quite modest. 
For instance, Gombosi et al. used the parameter $\tau$ to calculate 
the signal propagation speed in the telegraph equation by both methods. 
The correct derivation gave $v \sqrt{5/11}$, where $v$ is the particle speed, 
compared with $v \sqrt{1/3}$ when simple truncation was used---about 17\% difference. 
Effenberger and Litvinenko (2014) showed that the telegraph equation 
for focused transport, obtained by the simpler method, yields a solution that 
agrees well with the numerical solution of the original Fokker--Planck equation 
on an infinite interval. It is also worth stressing that the telegraph approximation 
was shown to reproduce the evolution of the particle density profile much more accurately 
than the diffusion approximation ($\tau=0$), especially when the focusing is strong 
(Litvinenko \& Noble 2013; Effenberger \& Litvinenko 2014). 

It may be useful to rewrite Equation (\ref{eq:FPE}) in terms of the linear density
\begin{equation}
  F(z, t) = \exp ( \xi z) F_0(z, t) ,
\label{eq:FandF0}
\end{equation}
which is the number of particles per line of force per unit distance
parallel to the mean magnetic field. Clearly, the linear density satisfies 
\begin{equation}
  \frac{\partial F}{\partial t} + 
  \tau \frac{\partial^2 F}{\partial t^2} = 
  \kappa_{\|} \frac{\partial^2 F}{\partial z^2} - 
  \xi \kappa_{\|} \frac{\partial F}{\partial z} . 
\label{eq:telF}
\end{equation}
The descriptions in terms of $F_0$ and $F$ are mathematically equivalent, 
and the choice of $F_0$ or $F$ is a matter of convenience (Earl 1981). 
For instance, particle conservation in the absence of sources and sinks 
is conveniently expressed as 
\begin{equation}
  N(t) = 2 \int F dz = \mbox{const} . 
\label{eq:Nconst}
\end{equation}
Note for clarity that Litvinenko \& Schlickeiser (2013) did not work with 
the linear density and they used the notation  $F(z, t)$ instead of $F_0(z, t)$.

\section{Boundary conditions}

Masoliver et al. (1992, 1993) obtained boundary conditions 
for the standard telegraph equation in the presence of reflecting 
or absorbing boundaries. We generalize the arguments in 
Masoliver et al. (1992, 1993) and derive the boundary conditions for 
a modified telegraph equation with a convective term 
that can describe the focused cosmic-ray transport. 

Decomposing the particle distribution function into two components 
corresponding to particles moving to the right, $a(z, t)$, 
and to the left, $b(z,t)$, we have the coupled equations 
that reduce to Equation (1) in Masoliver et al. (1992) in the limit $\xi=0$: 
\begin{equation}
  \partial_t a = 
  -\sqrt{\frac{\kappa_{\|}}{\tau}} \partial_z a + 
  \frac{1}{2 \tau} (b-a) + 
  \frac{1}{2} \xi \sqrt{\frac{\kappa_{\|}}{\tau}} (a+b) ,
\label{eq:a}
\end{equation}
\begin{equation}
  \partial_t b = 
  \sqrt{\frac{\kappa_{\|}}{\tau}} \partial_z b + 
  \frac{1}{2 \tau} (a-b) -  
  \frac{1}{2} \xi \sqrt{\frac{\kappa_{\|}}{\tau}} (a+b) . 
\label{eq:b}
\end{equation}
The linear density 
\begin{equation} 
  F(z,t) = a+b 
\end{equation} 
satisfies the focused telegraph equation (\ref{eq:telF}). 
Physically, $w = \sqrt{\kappa_{\|} / \tau}$ is a signal propagation speed 
in the telegraph equation, which does not depend on the first-order terms 
in the equation. The first term on the right in Equation (\ref{eq:a}) 
describes convective transport, the second term describes the particle change 
of direction, and the third term describes the focusing effect of a nonuniform 
background magnetic field. The interpretation is similar for Equation (\ref{eq:b}). 

Consider first a region $z_1 \le z \le z_2$ with absorbing boundaries 
at $z=z_1$ and $z=z_2$, so that 
\begin{equation}
  a(z_1,t) = 0 , \quad b(z_2,t) = 0 . 
\label{eq:absab}
\end{equation}
Subtracting Equation (\ref{eq:b}) from Equation (\ref{eq:a}) 
and using Equation (\ref{eq:absab}) yields the boundary conditions 
\begin{equation}
  \pm \sqrt{\kappa_{\|} \tau} (\partial_z F - \xi F) 
  = F + \tau \partial_t F ,
\label{eq:absF}
\end{equation}
where the plus (minus) sign corresponds to the left (right) boundary 
at $z=z_1$ ($z=z_2$). The absorbing boundary conditions for the density $F_0$ 
follow from Equations (\ref{eq:FandF0}) and (\ref{eq:absF}): 
\begin{equation}
  \pm \sqrt{\kappa_{\|} \tau} \partial_z F_0 
  = F_0 + \tau \partial_t F_0 . 
\label{eq:absF0}
\end{equation}
If we formally set $\tau = 0$, the telegraph approximation simplifies 
to the focused diffusion model, termed pseudo-diffusion by Earl (1981), 
and the absorbing boundary conditions above simplify to the familiar 
condition $F_0 = F = 0$ at an absorbing boundary. 

Now suppose that reflecting boundaries are present at 
$z=z_1$ and $z=z_2$, so that 
\begin{equation}
  a(z_1,t) = b(z_1,t), \quad a(z_2,t) = b(z_2,t) . 
\label{eq:refab}
\end{equation}
In this case, subtracting Equation (\ref{eq:b}) from Equation (\ref{eq:a}) 
and using Equation (\ref{eq:refab}) leads to the boundary conditions 
\begin{equation}
  \partial_z F - \xi F = 0 
\label{eq:refF}
\end{equation}
and 
\begin{equation}
  \partial_z F_0 = 0 
\label{eq:refF0}
\end{equation}
at $z=z_1$ and $z=z_2$, which are the same conditions as those for 
both the diffusion equation and the telegraph equation considered by 
Masoliver et al. (1993). In the case of a single reflecting boundary, 
Equation (\ref{eq:refF}) immediately follows on integrating Equation (\ref{eq:telF}) 
and using the particle conservation given by Equation (\ref{eq:Nconst}). 

Finally, we note that Marshak (1947) advocated the use of an escape 
condition for the particle flux 
\begin{equation}
  S = \frac{v}{2} \int_{-1}^{1} \mu g_0 d\mu 
\end{equation}
as a more accurate alternative to the conventional absorbing boundary 
condition. Marshak's condition is given by 
\begin{equation}
  S = \int_{-1}^{0} \mu v F_0 d\mu 
\end{equation}
at the left boundary and 
\begin{equation}
  S = \int_{0}^{1} \mu v F_0 d\mu 
\end{equation}
at the right boundary. 
For example, in the weak focusing limit of the telegraph approximation, 
the flux is given by Equation (13) in Litvinenko \& Schlickeiser (2013), 
which yields 
\begin{equation}
  -\kappa_{\|} \partial_z F_0 + \tau \kappa_{\|} \partial_{tz} F_0 
  = \pm \frac{1}{2} F_0 
\label{eq:escape}
\end{equation}
in our dimensionless units, where the minus (plus) sign corresponds to 
the left (right) boundary at $z=z_1$ ($z=z_2$). In the limit  $\tau = 0$, 
the condition reduces to the corresponding boundary condition of the diffusion 
model (Weinberg \& Wigner 1958). In practice, the presence of a second mixed 
derivative $\partial_{tz} F_0$ in Equation (\ref{eq:escape}) makes 
the boundary condition difficult to use.

\section{Analytical solutions}

\subsection{Infinite interval}

Analytical solutions of the focused diffusion model are commonly used 
to model the transport of solar energetic particles in interplanetary space 
(e.g., Artmann et al. 2011 and references therein). 
Since the telegraph approximation should give a more accurate description 
of the cosmic-ray transport, analytical solutions of the telegraph approximation 
could be used to quantify the accuracy of the diffusion approximation 
or to validate the accuracy of a numerical solution of the Fokker--Planck 
equation for an evolving particle distribution function.
Here we illustrate how initial and boundary value problems for 
the modified telegraph equation can be solved by Laplace transform for infinite 
and semi-infinite intervals and by Fourier series for a finite interval.

For simplicity we assume a symmetric point source at $z=z_0$, so that 
$a(z,0) = b(z,0)$, and the initial conditions are given by 
\begin{equation}
  F_0 (z, 0) = \delta (z-z_0) , \quad \partial_t F_0 (z, 0) = 0 , 
\label{eq:initial}
\end{equation}
where the second condition follows from Equations (\ref{eq:a}) and (\ref{eq:b}). 
The telegraph equation is linear, and so we omit the normalization constant 
$\frac{1}{2} \exp(-\xi z_0) N$ in $F_0 (z, 0)$ for brevity. 

Consider first the initial value problem for an infinite interval. Although 
in this case the solution can be obtained by a change of variables, 
which reduces Equation (\ref{eq:telF0}) to the standard telegraph equation 
(e.g., Kevorkian 2000), it is instructive to solve the problem 
directly using the Laplace transform. 
The transform of Equation (\ref{eq:telF0}) reads 
\begin{equation}
  {\tilde F}_0^{\prime \prime} + \xi {\tilde F}_0^{\prime} - 
  \frac{1}{\kappa_{\|}} s (1 + \tau s) {\tilde F}_0 = 
  - \frac{1}{\kappa_{\|}} (1 + \tau s) \delta (z-z_0) , 
\label{eq:telF0trans}
\end{equation}
with the Laplace transform ${\tilde F}_0 (z,s) = L[F_0(z,t)]$ 
and the prime denotes differentiation with respect to $z$. 
Solving the ordinary differential equation yields the transform 
\begin{equation}
  {\tilde F}_0 (z,s) = 
  \frac{1 + \tau s}{2 \kappa_{\|} \eta} 
  \exp \left( -\frac{\xi}{2} (z-z_0) - \eta |z-z_0| \right) , 
\end{equation}
where 
\begin{equation}
  \eta = \sqrt{\frac{\xi^2}{4} + \frac{s(1 + \tau s)}{\kappa_{\|}}} . 
\end{equation}
Consequently the solution of the initial value problem is given by 
\begin{equation}
  F_0(z, t) = G_0 + \tau \partial_t G_0 , 
\label{eq:F0=G0tauG0}
\end{equation}
where  
\begin{equation}
  G_0(z, t) = L^{-1} \left[ \frac{1}{2 \kappa_{\|} \eta} 
  \exp \left( -\frac{\xi}{2} (z-z_0) - \eta |z-z_0| \right) \right] 
\end{equation}
is the fundamental solution of the modified telegraph equation. 
Evaluating the inverse transform yields 
\begin{equation}
  G_0 (z, t) = \frac{1}{2 \sqrt{\kappa_{\|} \tau}} 
  \exp \left( -\frac{\xi}{2} (z-z_0) - \frac{t}{2 \tau} \right) 
  I_0 (u) H \left( t - \sqrt{\frac{\tau}{\kappa_{\|}}} |z-z_0| \right) . 
\label{eq:G0}
\end{equation}
Here, $H$ is the Heaviside step function, 
$I_0$ is a modified Bessel function, and its argument is 
\begin{equation}
  u = \frac{1}{2} 
  \sqrt{ \left( 1 - \xi^2 \kappa_{\|} \tau \right) 
  \left( \frac{t^2}{\tau^2} - \frac{(z-z_0)^2}{\kappa_{\|} \tau} \right) } . 
\end{equation}
Note that we assume $ \xi^2 \kappa_{\|} \tau <  1 $. 
The inequality was shown to be valid both in the case of weak focusing 
and anisotropic scattering (Litvinenko \& Schlickeiser 2013) and in the case 
of strong focusing and isotropic scattering (Litvinenko \& Noble 2013). 
For instance, $( 1 - \xi^2 \kappa_{\|} \tau ) / \tau = 1$ for isotropic scattering. 
Substituting Equation (\ref{eq:G0}) into Equation (\ref{eq:F0=G0tauG0}) 
leads to a solution in terms of the modified Bessel functions $I_0$ and $I_1$: 
\begin{eqnarray}
  F_0 (z, t) = \frac{1}{4 \sqrt{\kappa_{\|} \tau}}  
  \exp \left( -\frac{\xi}{2} (z-z_0) - \frac{t}{2 \tau} \right) \times 
\nonumber
\\
  \left[ I_0 (u) + \left( 1 - \xi^2 \kappa_{\|} \tau \right) 
  \frac{t}{2 \tau} \frac{I_1(u)}{u} \right] 
  H \left( t - \sqrt{\frac{\tau}{\kappa_{\|}}} |z-z_0| \right) 
\nonumber
\\
  + \frac{1}{2} \exp \left( -\frac{\xi}{2} (z-z_0) - \frac{t}{2 \tau} \right) 
  \left[ \delta \left( \sqrt{\frac{\kappa_{\|}}{\tau}} t - (z-z_0) \right) + 
  \delta \left( \sqrt{\frac{\kappa_{\|}}{\tau}} t + (z-z_0) \right) \right] , 
\label{eq:F0-full}
\end{eqnarray}
which slightly generalizes Equations (18) and (19) 
in Effenberger \& Litvinenko (2014) who compared the solution with 
both the prediction of the diffusion model and a numerical solution 
of the original Fokker--Planck equation. 

\subsection{Semi-infinite interval: a reflecting boundary}

 Now consider the initial value problem, specified by Equations
(\ref{eq:refF0}) 
 and (\ref{eq:initial}), on a semi-infinite
interval $z>0$. Physically, 
 a reflecting inner boundary condition
at $z_1 = 0$ may correspond to the transport of solar energetic
particles, accelerated close to the solar surface. The reflection of
particles traveling towards the sun is caused by a magnetic bottle
effect of the strongly converging magnetic field. In this case
the solution of the transformed Equation (\ref{eq:telF0trans}), 
which satisfies $\partial_z {\tilde F}_0 (0,s) = 0$, is given by 
\begin{eqnarray}
  {\tilde F}_0 (z,s) = 
  \frac{1 + \tau s}{\kappa_{\|} \eta (2 \eta + \xi)} 
  \exp \left( -\frac{\xi}{2} (z-z_0) \right) \times 
\nonumber
\\
  \left[ \left( \eta + \frac{1}{2} \xi \right) 
  \exp \left( - \eta |z-z_0| \right) + 
  \left( \eta - \frac{1}{2} \xi \right) 
  \exp \left( - \eta (z+z_0) \right) \right] . 
\end{eqnarray}

As an interesting aside, note that the mean age $T$ of particles 
at a given location $z$ can be elegantly expressed in terms of 
the Laplace transform ${\tilde F}_0$: 
\begin{equation}
  T = \frac{\int_0^{\infty} t F_0 dt}{\int_0^{\infty} F_0 dt} 
  = - \partial_s {\tilde F}_0 (z,0) . 
\end{equation}
For example, suppose that a particle source is located very close 
to the solar surface, so that we may take $z_0 = 0$. 
We calculate 
\begin{equation}
  T = \frac{1}{\kappa_{\|} \xi^2} + 
  \frac{z}{\kappa_{\|} \xi} - \tau , 
\end{equation}
which agrees with the advection-dominated limit of 
Equation (4) in Jokipii (1976).

The solution of the initial value problem is again given by 
Equation (\ref{eq:F0=G0tauG0}), but now the fundamental solution 
is as follows: 
\begin{equation}
  G_0(z, t) = L^{-1} \left[ 
  \frac{1}{2 \kappa_{\|} \eta} 
  \exp \left( -\frac{\xi}{2} (z-z_0) \right) 
  \left[ \exp \left( - \eta |z-z_0| \right) + 
  \frac{2 \eta - \xi}{2 \eta + \xi} 
  \exp \left( - \eta (z+z_0) \right) \right] \right] . 
\end{equation}
For simplicity, consider again a particle source located near the boundary. 
Setting $z_0 = 0$ and keeping in mind that $z>0$, we have 
\begin{equation}
  G_0(z, t) = L^{-1} \left[ 
  \frac{2}{\kappa_{\|} (2 \eta + \xi)} 
  \exp \left( - \frac{1}{2} ( 2 \eta + \xi ) z  \right) \right] . 
\end{equation}
The inverse Laplace transform can be expressed in terms of 
a table transform (Erd\'elyi et al. 1954) by noticing that 
\begin{equation}
  \eta^2 = \frac{\tau}{\kappa_{\|}} \left[ 
  \left( s + \frac{1}{2 \tau} \right)^2 - \alpha^2 \right] 
\end{equation}
where 
\begin{equation}
  \alpha^2 = \frac{1 - \xi^2 \kappa_{\|} \tau}{4 \tau^2} . 
\end{equation}
The resulting fundamental solution is as follows: 
\begin{eqnarray}
  G_0 (z, t) = \frac{1}{\sqrt{\kappa_{\|} \tau}} 
  \exp \left( - \frac{t}{2 \tau} \right) 
  H \left( t - \sqrt{\frac{\tau}{\kappa_{\|}}} z \right) \times 
\nonumber
\\
  \left[ 
  \exp \left( - \frac{\xi}{2} \sqrt{\frac{\kappa_{\|}}{\tau}} t \right) 
  + \alpha \int_0^{\sqrt{t^2 - \tau z^2 / \kappa_{\|}}} 
  \exp \left( - \frac{\xi}{2} \sqrt{\frac{\kappa_{\|}}{\tau}} 
  \sqrt{t^2-w^2} \right) I_1 (\alpha w) dw \right] . 
\end{eqnarray}

The result can be verified in the limiting case of a uniform 
magnetic field ($\xi = 0$) when the solution simplifies to 
\begin{equation}
  G_0 (z, t) = \frac{1}{\sqrt{\kappa_{\|} \tau}} 
  \exp \left( - \frac{t}{2 \tau} \right) 
  I_0 (u_0) H \left( t - \sqrt{\frac{\tau}{\kappa_{\|}}} z \right) 
\end{equation}
where 
\begin{equation}
  u_0 = \frac{1}{2} 
  \sqrt{ \frac{t^2}{\tau^2} - \frac{z^2}{\kappa_{\|} \tau} } . 
\end{equation}
Consequently, 
\begin{eqnarray}
  F_0 (z, t) = \frac{1}{2 \sqrt{\kappa_{\|} \tau}}  
  \exp \left( - \frac{t}{2 \tau} \right) 
  \left[ I_0 (u_0) + \frac{t}{2 \tau} \frac{I_1(u_0)}{u_0} \right] 
  H \left( t - \sqrt{\frac{\tau}{\kappa_{\|}}} z \right) 
\nonumber
\\
  + \exp \left( - \frac{t}{2 \tau} \right) 
  \delta \left( \sqrt{\frac{\kappa_{\|}}{\tau}} t - z \right) . 
\end{eqnarray}
In other words, the solution on the semi-infinite interval $z>0$ 
with a source at the edge of the interval in the absence of focusing 
is simply double the solution of the corresponding initial value problem 
for an infinite interval. The result immediately follows on applying 
the method of images to the problem. 

Finally, for a uniform magnetic field, the solution on a semi-infinite
interval with an absorbing boundary is given by Equation (19) in
Masoliver et al. (1992).

\subsection{Finite interval: reflecting boundaries}
The relatively simple form of the reflecting boundary conditions 
makes it possible to express the solution of the telegraph equation 
on a finite interval in terms of a Fourier series. 
The series solution is particularly convenient for long times, 
when the first few terms of the series accurately approximate the solution. 
Suppose that two reflecting boundaries are present at $z_1 = 0$ and $z_2 = l$, 
$F_0 (z, 0)$ is given for $0<z<l$, and $\partial_t F_0 (z, 0) = 0$ for simplicity. 
Straightforward application of the method of separation of variables to 
Equations (\ref{eq:telF0}) and (\ref{eq:refF0}) leads to 
\begin{equation}
  F_0 (z, t) = c_0 + \sum_{n=1}^{\infty} c_n Z_n(z) T_n(t) , 
\label{eq:reflect}
\end{equation}
where 
\begin{equation}
  Z_n (z) = 
  \exp \left( - \frac{\xi z}{2} \right) 
  \left[ \cos \left( \frac{\pi n z}{l} \right) + 
  \frac{\xi l}{2 \pi n} \sin \left( \frac{\pi n z}{l} \right) \right] , 
\end{equation}
\begin{equation}
  T_n (t) = 
  \exp \left( - \frac{t}{2 \tau} \right) 
  \left[ \cosh \omega_n t + \frac{1}{2 \omega_n \tau} \sinh \omega_n t \right] , 
\end{equation}
\begin{equation}
  \omega_n = \left[ 
  \frac{1}{4 \tau^2} - \frac{\kappa_{\|}}{\tau} 
  \left( \frac{\xi^2}{4} + \left( \frac{\pi n}{l} \right)^2 \right)
  \right]^{1/2} . 
\end{equation}
The initial conditions yield 
\begin{equation}
  c_0 = 
  \frac{\xi}{[\exp(\xi l) - 1]} 
  \int_0^l \exp \left( \xi z \right) F_0 (z, 0) dz , 
\end{equation}
\begin{equation}
  c_n = 
  \frac{2}{l} 
  \left[ 1 + \left( \frac{\xi l}{2 \pi n} \right)^2 \right]^{-1}
  \int_0^l \exp \left( \frac{\xi z}{2} \right) 
  \left[ \cos \left( \frac{\pi n z}{l} \right) + 
  \frac{\xi l}{2 \pi n} \sin \left( \frac{\pi n z}{l} \right) \right] F_0 (z, 0) dz . 
\end{equation}
In the limit $\xi = 0$, the solution for the initial profile 
$F_0 (z, 0) = \delta (z-z_0)$ agrees with Equation (19) in Masoliver et al. (1993). 

\subsection{Finite interval: absorbing boundaries}
\label{sec:parabolic}
The appearance of two partial derivatives in Equation (\ref{eq:absF0}) 
complicates the solution of the telegraph equation with absorbing boundaries. 
In the case $\xi = 0$ and $F_0 (z, 0) = \delta (z-z_0)$, 
Equation (25) in Masoliver et al. (1992) gives an exact solution 
in terms of an infinite series of Bessel functions. 
Since numerical solutions might not capture important qualitative 
features of the solution, it is natural to seek an approximate analytical solution. 
Here, we illustrate an integral approximation method, 
similar to the heat-balance approximation in heat conduction 
(e.g., Crank 1984; Hill \& Dewynne 1987). 

Consider a finite region with absorbing boundaries 
at $z_1 = 0$ and $z_2 = l$. The boundary value problem for Equation (\ref{eq:telF0}) 
might serve as the basis for a simple model for the escape of galactic cosmic rays 
away from the galactic plane (Schlickeiser 2009). 
The idea of the approximation is that, 
instead of solving Equation (\ref{eq:telF0}) exactly, 
we integrate it with respect to $z$ from $0$ to $l$ and seek a solution 
of a specified functional form that satisfies the boundary conditions. 
As a simple illustration, consider 
\begin{equation}
  F_0(z, t) \approx f_1 (z) f_2 (t) 
\label{eq:parabolic}
\end{equation}
and require that it satisfies both the absorbing boundary conditions, 
given by Equation (\ref{eq:absF0}), and an ordinary differential equation for $f_2 (t)$, 
obtained by integrating Equation (\ref{eq:telF0}) over $z$. 
On choosing a parabolic density profile 
\begin{equation}
  f_1 (z) = 1 + kz(l-z) , 
\label{eq:parabolic-profile}
\end{equation}
it is straightforward to verify that 
\begin{equation}
  \tau \ddot{f_2} + \dot{f_2} + 
  \frac{2 \kappa_{\|} k}{1+kl^2/6} f_2 = 0 , 
\label{eq:f2ODE}
\end{equation}
and so a possible solution is given by 
\begin{equation}
  f_2 (t) = \mbox{const} \exp (\lambda t) , 
\end{equation}
where the constants $k$ and  $\lambda$ follow from 
Equation (\ref{eq:f2ODE}) and the boundary conditions: 
\begin{equation}
  k = \frac{1 + \lambda \tau}{\sqrt{\kappa_{\|} \tau} l} , 
\label{eq:parabolic-k}
\end{equation}
\begin{equation}
  \lambda = -\frac{2}{l} \sqrt{\frac{\kappa_{\|}}{\tau}} \frac{1}{1+kl^2/6} . 
\end{equation}
It follows that 
\begin{equation}
  \lambda = - \frac{1}{2 \tau} - \frac{3}{l} \sqrt{\frac{\kappa_{\|}}{\tau}} 
  \pm \sqrt{ 
  \left( \frac{1}{2 \tau} + \frac{3}{l} \sqrt{\frac{\kappa_{\|}}{\tau}} \right)^2 
  - \frac{12 \kappa_{\|}}{\tau l^2} } , 
\label{eq:parabolic-lambda}
\end{equation}
and so the approximate solution describes the evolution of an initially broad 
(parabolic) particle density profile. 

\section{Stochastic simulations of the Fokker--Planck equation}

\subsection{Numerical method}

To quantitatively assess the accuracy of the telegraph approximation
in the presence of boundaries, we use numerical solutions 
of the corresponding Fokker--Planck equation. 
The numerical approach is based on solutions to an
equivalent system of stochastic differential equations (SDEs), similar
to the method employed in Litvinenko \& Noble (2013) and Effenberger
\& Litvinenko (2014). In the following, we use the Milstein
approximation scheme, given by Equations (29) and (30) in Effenberger \&
Litvinenko (2014), and we refer the reader to this paper for more
details on the numerics (see also Kopp et al. 2012, and references
therein for more details on the SDE method).

In the present study, we have to take special care of the boundary
conditions required for the Fokker--Planck equation. In practice, this
means that the trajectories of pseudo-particles are integrated according 
to their stochastic evolution equations, until they cross a boundary (say at
$z=0$ or $z=l$). At this point, either the particle speed is reversed 
(and so its pitch angle changes its sign at a reflecting
boundary) or the particle is discarded for the rest of the simulation 
at an absorbing boundary. The final distribution function is
constructed in the usual way, by applying an appropriate binning
procedure to the particle positions, and normalizing the result for
comparison with the analytic predictions.

We consider the case of isotropic scattering, for which the
Fokker--Planck pitch-angle scattering coefficient $D_{\mu\mu}$ is given
by
\begin{equation}
D_{\mu\mu}=D_0(1-\mu^2).
\label{eq:Dmumu}
\end{equation}
In this case, the coefficients in the telegraph equation
(\ref{eq:telF0}) are given by 
\begin{eqnarray}
\kappa_\parallel &=& \frac{\coth\xi}{\xi} - \frac{1}{\xi^2}, \\
\tau            &=& \frac{\tanh\xi}{\xi} 
\label{eq:kappaandtau}
\end{eqnarray}
(Litvinenko \& Noble 2013; He \& Schlickeiser 2014, and references therein). 
We have $\kappa_\parallel \approx 1/3$ and $\tau \approx 1$ 
in the weak focusing limit $\xi^2 \ll 1$. 

\subsection{Semi-infinite interval: a reflecting boundary}
\begin{figure}[ht]
\noindent\centering\includegraphics[width=0.49\textwidth]{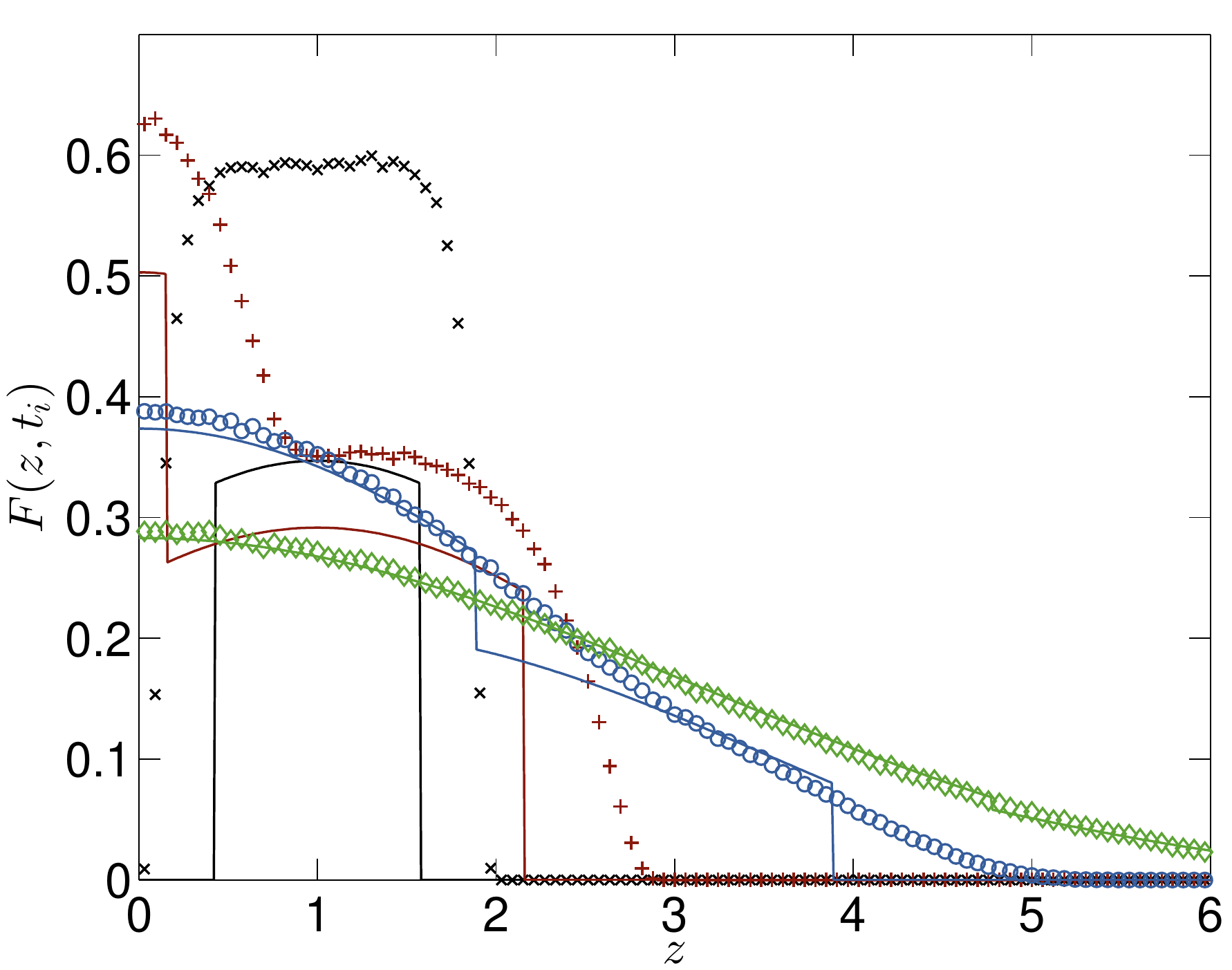}
\caption{Isotropic density $F(z, t_i)$ at four different times $t_1=1$
  (black, 'x'), $t_2=2$ (red, '+'), $t_3=5$ (blue, 'o') and $t_4=10$
  (green, '$\diamond$') for the semi-infinite domain with reflecting
  boundary. The solid lines show the solution of the telegraph
  equation, given by Equation (\ref{eq:semireflect}). The symbols show the
  numerical results, obtained from $10^6$ particles starting at
  $z_0=1.0$, and averaging without regard to the pitch angle of the
  particles.}
\label{fig:semiinf}
\end{figure}
We first consider the positive semi-infinite region with a reflecting
boundary at $z=0$. Particles are injected at time $t_0=0$ and position
$z_0=1$ with isotropic pitch angle. Masoliver et al. (1993) gave a
solution for the telegraph equation for a case without focusing based
on the method of images (their Equations (14) and (16)). Their
solution follows from our Equation (\ref{eq:F0-full}) in the limit 
$\xi\rightarrow 0$ when we apply the method of images: 
\begin{equation}
  F^{\mathsf{refl}} (z,t) = F_0(z,t\vert{z_0}) + F_0(z,t\vert{-z_0}) \, ,
 \label{eq:semireflect}
\end{equation}
where $F_0(z,t\vert{z_0})$ denotes the solution with injection at
$z_0$. Note for clarity that in Masoliver et al. (1993) the signal
propagation speed is denoted as $c$ and their $T$ is equal to our
$\tau$. The method of images is not applicable in the case $\xi \neq 0$.

We concentrate on the case $\xi=0$ to describe the effect of the
boundaries on the particle distribution. We compare the analytical
solution for the semi-infinite domain with simulation results for the
Fokker--Planck equation as described above. Figure~\ref{fig:semiinf}
shows the evolution of the particle distribution at four different
times, calculated from an injection of $10^6$ pseudo-particles. The
comparison shows a good agreement after a few scattering times, 
as was the case for the solution on an infinite interval 
(Effenberger \& Litvinenko 2014). The disagreement between 
the numerical and approximate analytical solutions, however, is significant 
at earlier times. The overall amplitude of the distribution appears to be
underestimated by the telegraph solution.  Note that, while the
telegraph solution conserves the particle number, the
$\delta$-functional contributions are omitted in
Figure~\ref{fig:semiinf}.  Furthermore, the slower signal propagation
speed, when compared to the particle speed, results in positions of
the discontinuities, which underestimate the fast particle spread
beyond the limits imposed by the signal propagation speed, especially
at early times. At times $t=5$ and $t=10$ the agreement is
significantly better, and the discontinuities have almost disappeared. 

\begin{figure}[ht]
\noindent\centering\includegraphics[width=0.49\textwidth]{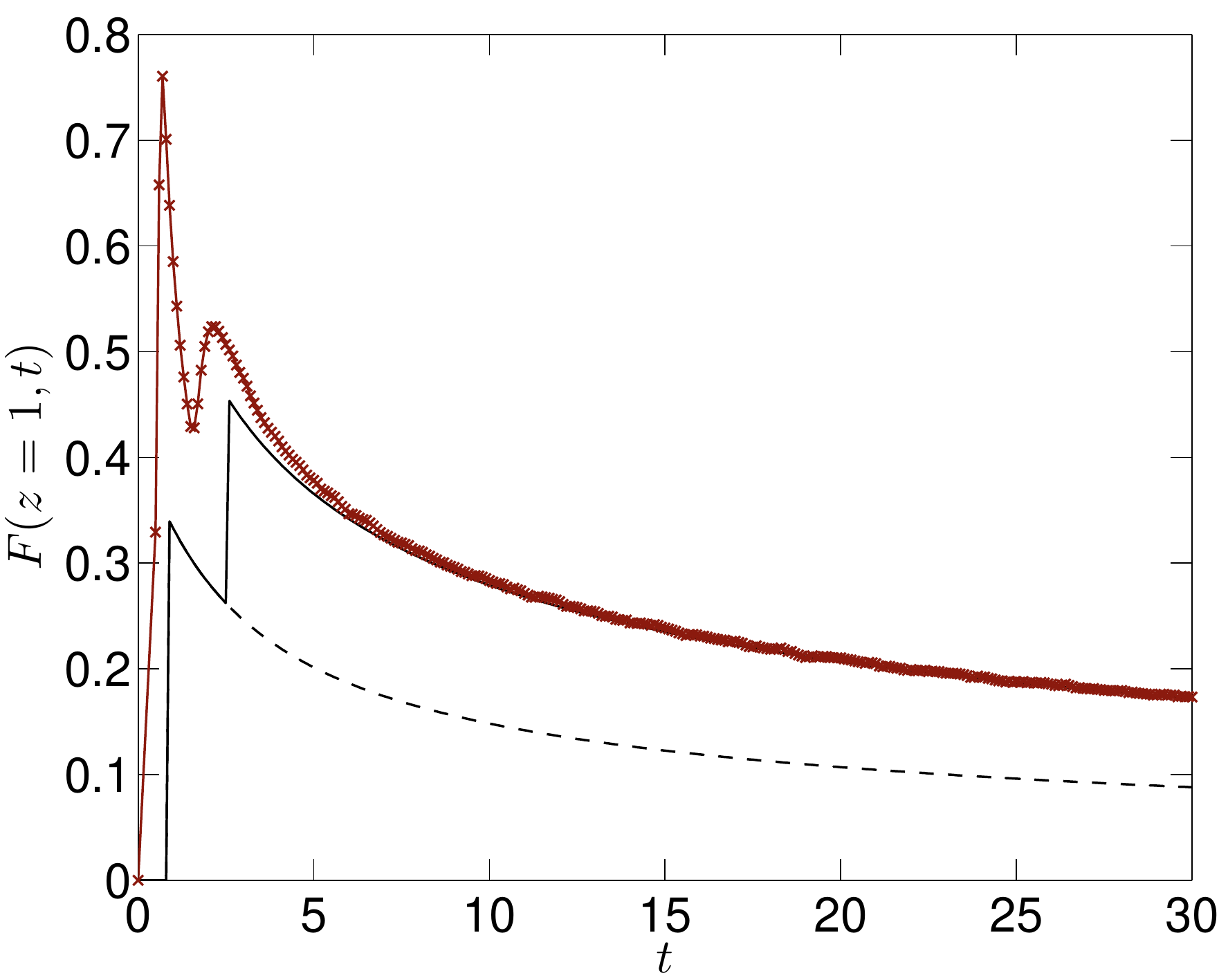}
\caption{Time profile at $z=1$ for the semi-infinite interval with
a reflecting left boundary. Particles are injected at $z_0=0.5$ to
illustrate the reflection effect and the resulting discontinuities in
the telegraph solution. The solid black line gives the solution of
Equation~(\ref{eq:semireflect}), while the dashed line gives the
solution for an infinite interval (without a reflecting boundary at
the origin). The symbols connected with a red line show the numerical
SDE solution obtained from the injection of $10^6$ particles.}
\label{fig:semiinf-time}
\end{figure}
In addition to the spatial density profiles, time profiles of particle intensities 
can provide an important tool for analyzing solar energetic particle data. 
Figure~\ref{fig:semiinf-time} illustrates the effect of a reflecting boundary at $z=0$, 
say due to the magnetic mirroring of energetic particles accelerated in the solar corona. 
To emphasize the effect of the reflecting boundary for illustrative purposes, we assume that 
the particle injection occurs at $z_0 = 0.5$ and plot the time profile at $z=1$. 
A jump in the particle intensity caused by the arrival of reflected particles 
is clearly visible. The early arrival of non-scattered particles at dimensionless time 
$t = (z-z_0)/v = 0.5$ causes the discrepancy between the numerical 
Fokker--Planck solution and the telegraph approximation, which almost disappears 
when the diffusing reflected particles start arriving at time $t = (z+z_0)/w \approx 2.6$ 
and decreases even further as the transport becomes more diffusive for later times.

\subsection{Semi-infinite interval: an absorbing boundary}
\begin{figure}[ht]
\noindent\centering\includegraphics[width=0.49\textwidth]{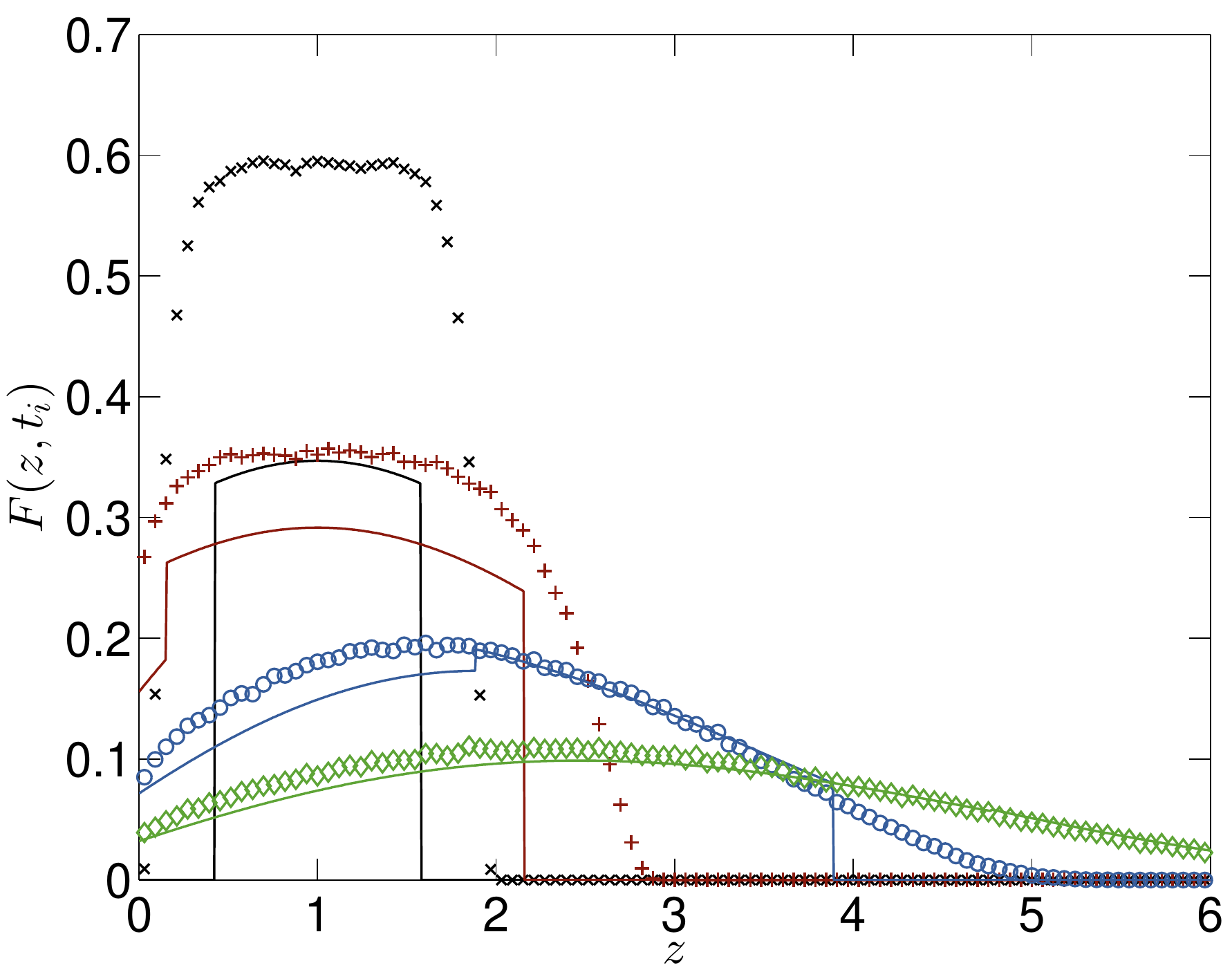}
\caption{Similar to Figure \ref{fig:semiinf}, but now with an
  absorbing left boundary condition.}
\label{fig:semiinf-refl}
\end{figure}
We consider a setting similar to the one in Section 5.2, with
initial particle injection again at $z_0=1$, but now for a left-hand
side absorbing boundary condition on the semi-infinite interval.
Figure~\ref{fig:semiinf-refl} compares the analytic result for the
case without focusing (Equation (19) in Masoliver et al. 1992) with the
numerical Fokker--Planck results for $10^6$ particles. For the purpose 
of comparison, the solution of Masoliver needs to be rescaled with $z=cy$ 
and normalized with the signal speed, due to the different nondimensionalization.

As in the previous case with reflecting boundary conditions, 
we find that the agreement of the analytical and numerical solutions 
improves as later times. A noteworthy feature is the
non-vanishing boundary values at an absorbing boundary, in contrast
to the boundary condition $F = F_0 = 0$ in the diffusion approximation
(see also Section \ref{sec:finite-absorb} below).

\subsection{Finite interval: reflecting boundaries}
\begin{figure}[ht]
\includegraphics[width=0.49\textwidth]{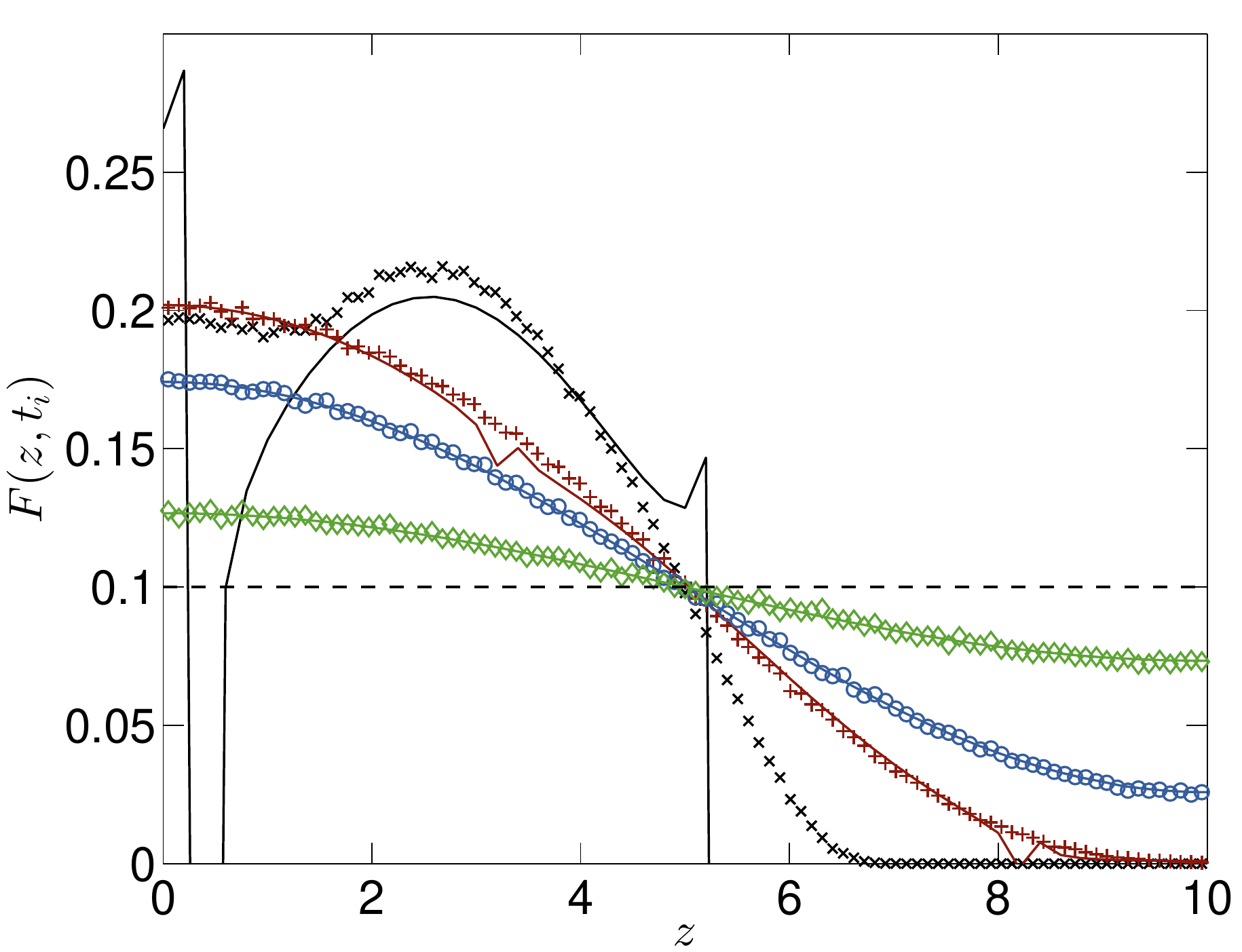}
\includegraphics[width=0.49\textwidth]{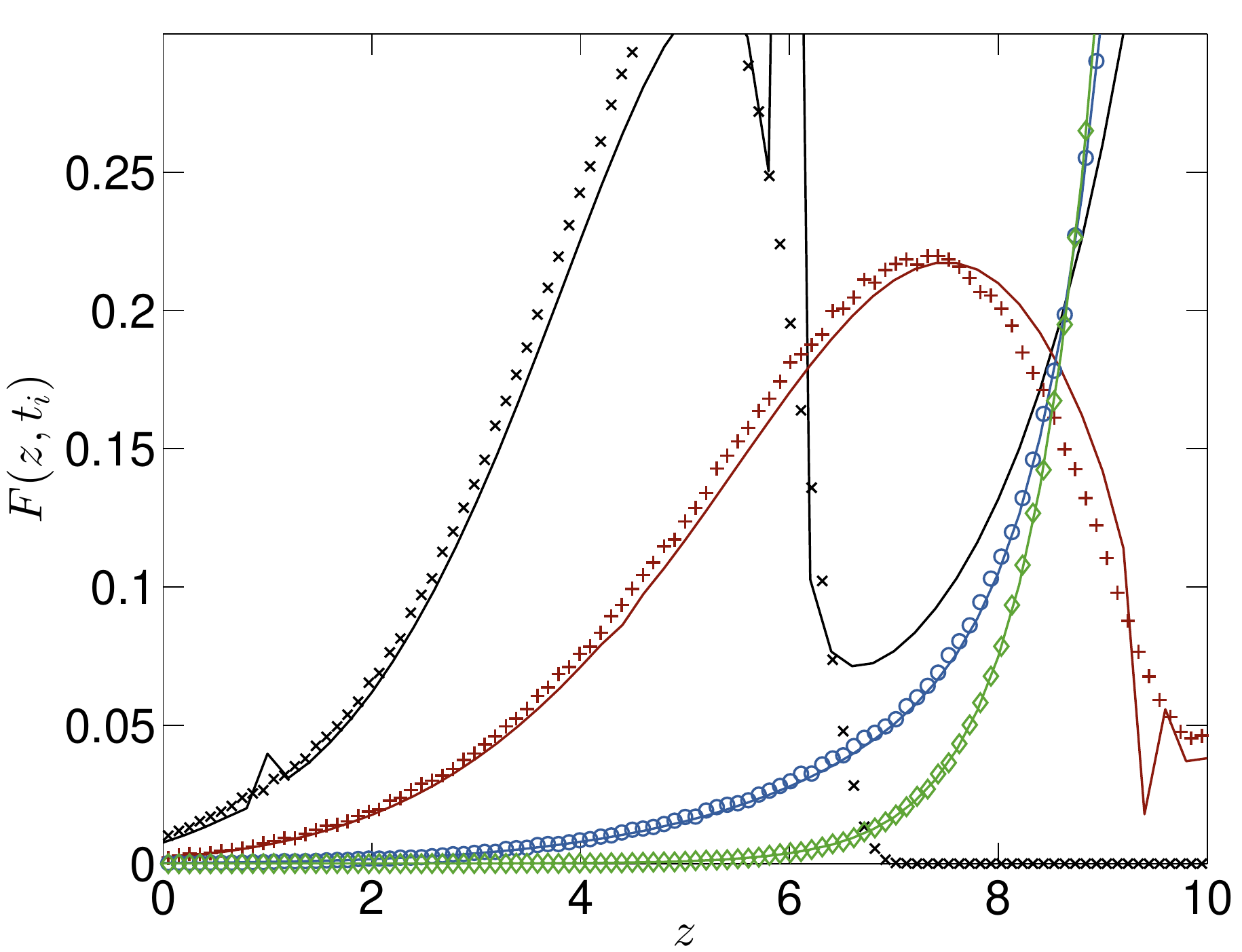}
\caption{No focusing ($\xi=0$, left panel) and strong focusing
  ($\xi=1.5$, right panel) results for the isotropic linear density
  $F(z, t_i)$ on a finite domain of length $l=10$ with injection at
  $z_0=2.5$ and reflecting boundaries. The four different times
  displayed are $t_1=5$ (black, 'x'), $t_2=10$ (red, '+'), $t_3=20$
  (blue, 'o') and $t_4=50$ (green, '$\diamond$'). The solid lines show the
  solution of the telegraph equation, given by Equation
  (\ref{eq:reflect}). The symbols show the numerical results, obtained
  from $10^6$ particles starting at $z_0$. In the left panel, the
  dashed line indicates the constant steady-state solution.}
\label{fig:tworefl}
\end{figure}

As the next example, we consider a finite interval with reflecting
boundaries. We evaluate equation (\ref{eq:reflect}) with $N=1000$
terms in the Fourier series on a finite domain of length $l=10$ with
an isotropic injection of particles at
$z_0=2.5$. Figure~\ref{fig:tworefl} shows a comparison between this
result and a stochastic simulation of $10^6$ particles giving a
solution to the corresponding Fokker--Planck Equation
(\ref{eq:FPE}). We consider the cases of no focusing ($\xi=0$) and
strong focusing ($\xi=1.5$).

At relatively early times, say $t_1=5$, the Fourier series solution
shows limited applicability near the discontinuities in the solution
of the telegraph equation. Similar to the previous cases of infinite
and semi-infinite domain, the Fokker--Planck solution exhibits a
relatively smooth profile at the discontinuities, and at later stages
the agreement is very good, both for weak and strong focusing. 
For long times, the solution approaches the steady state solution, 
given by the term $c_0$ in Equation (\ref{eq:reflect}): 
\begin{equation}
  F_0 (z, t \to \infty) \to c_0 = \frac{\xi N}{2 (\exp(\xi l) - 1)} , 
\end{equation}
which reduces to $F_0 = N/2l$ in the limit $\xi = 0$.

\subsection{Finite interval: absorbing boundaries}
\label{sec:finite-absorb}
\begin{figure}[ht]
\includegraphics[width=0.49\textwidth]{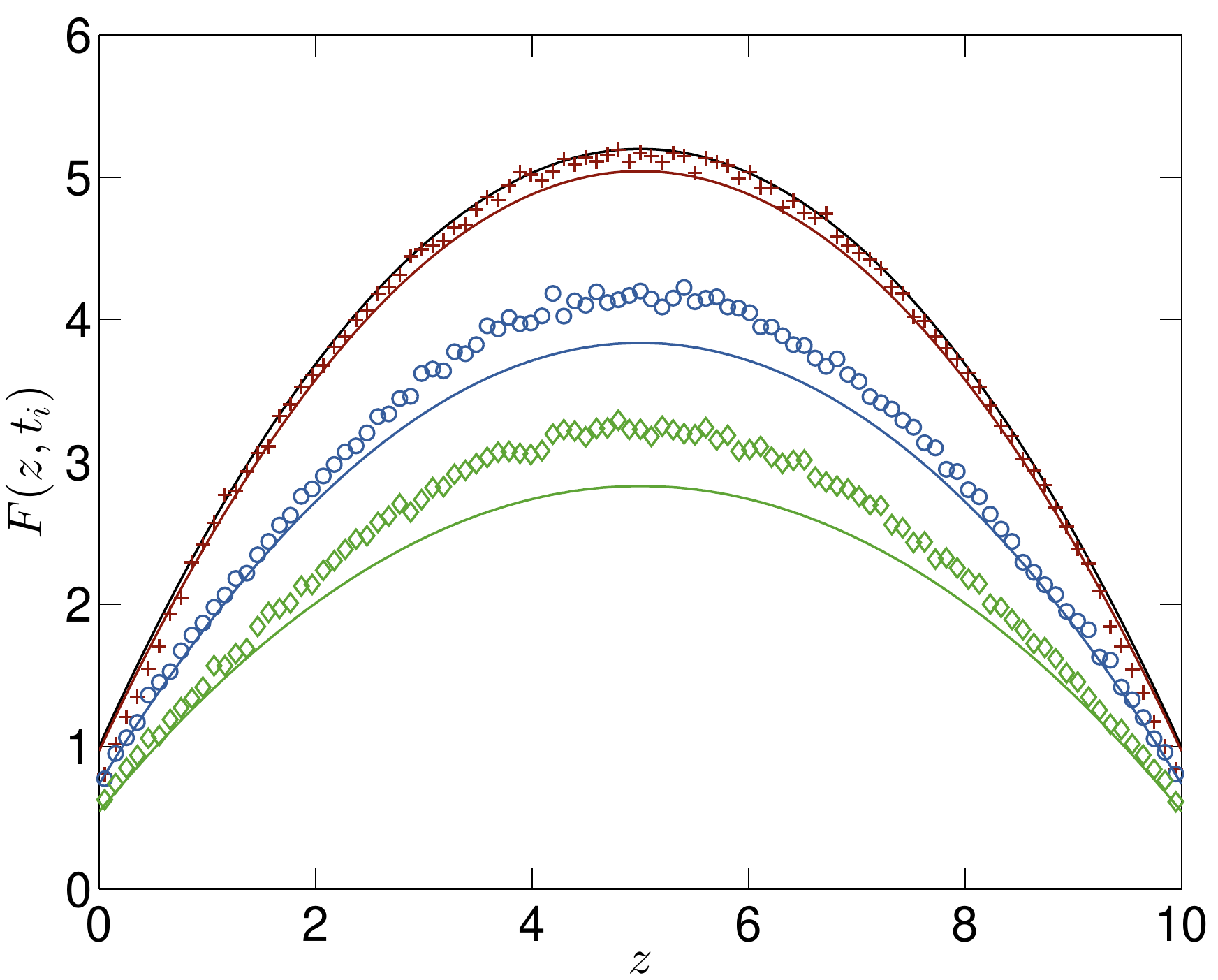}
\includegraphics[width=0.50\textwidth]{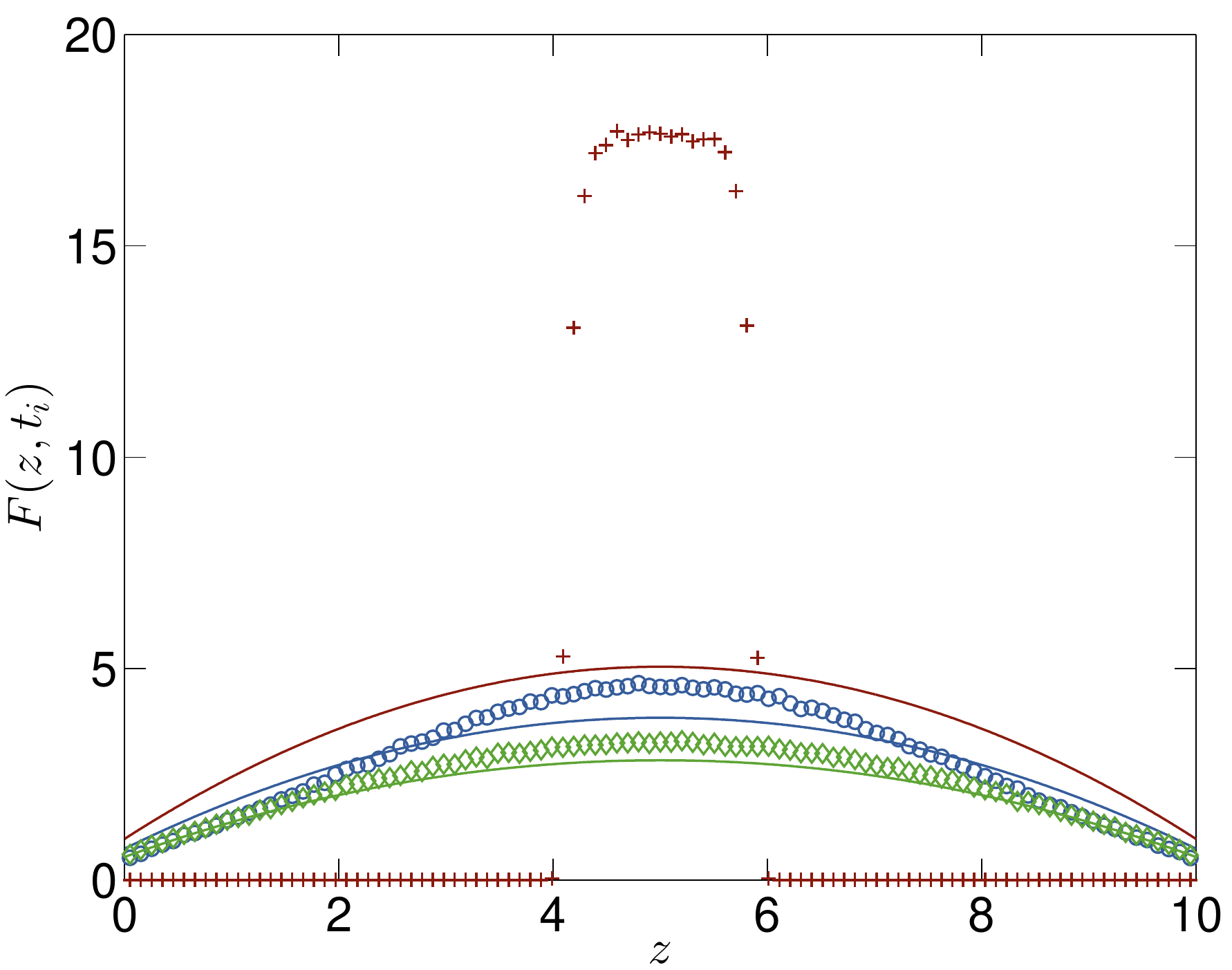}
\caption{Comparison of the parabolic solution for the finite interval
  with absorbing boundaries (Equation (\ref{eq:parabolic}), solid lines)
  to a SDE solution (symbols) of the Fokker--Planck equation with
  $10^6$ particles initialized with the parabolic profile at $t=0$
  (solid black line) with a rejection sampling method (left panel) or
  with a point injection at $z_0=5$ (right panel). The three different times
  displayed are $t_1=1$ (red, '+'), $t_2=10$ (blue, 'o') and $t_3=20$
  (green, '$\diamond$').}
\label{fig:twoabsorb}
\end{figure}
Finally, we consider a finite domain of length
$l=10$ with two absorbing boundaries. We compare the approximate parabolic
solution of section \ref{sec:parabolic} to SDE solutions of the
Fokker--Planck equation, initialized either directly with the parabolic
profile of equation (\ref{eq:parabolic-profile}) or with a point
injection at $z_0=5$. The values for $k$ and $\lambda$ are determined
from Equations (\ref{eq:parabolic-k}) and
(\ref{eq:parabolic-lambda}) as
\begin{eqnarray}
\lambda &=& -0.0304 \\
k &=& 0.1679 \, .
\end{eqnarray}
The second possible value for $\lambda$ would result in a negative
value for $k$ giving an inverted parabolic profile.

Figure~\ref{fig:twoabsorb} shows the two cases (parabolic initial
profile, left panel; delta-functional injection, right panel). For the left panel
plot, the solutions have been normalized to agree for the initial
profile. For the right panel, the solution is normalized to agree
approximately at time $t_3=20$. It is clear that the parabolic
solution captures both the general structure of the Fokker--Planck solution
and its approximate exponential decay behavior quite well. The
detailed time evolution and spatial dependence, however, are more
complicated, and for the delta-functional injection, the parabolic
profile only develops at a later stage. As mentioned before, in the
context of the semi-infinite interval, the telegraph approximation
allows for non-vanishing boundary values, in contrast to the
boundary condition $F = F_0 = 0$ in the diffusion approximation. This
appears to yield good agreement of the telegraph solution with the
numerical Fokker--Planck results in the vicinity of the absorbing
boundaries.

We could not confirm numerically the presence of discontinuities in the
analytical solution of Masoliver et al.\ (1992) for the initial isotropic
pitch-angle distribution of particles (see Dunkel et al. 2007, 
for discussion of this feature of the telegraph approximation, 
as well as for an alternative approach to diffusive transport modeling). 
An initial condition corresponding to two oppositely-traveling particle beams 
might give a better agreement of the numerical Fokker--Planck solution 
and the telegraph approximation. 

\section{Discussion}

In this paper we systematically developed the telegraph approximation
for particle transport in the presence of boundaries, taking into
account the adiabatic focusing effect in a non-uniform mean magnetic
field, which leads to coherent particle streaming along
the field. We derived reflecting and absorbing boundary conditions 
for a modified telegraph equation with a convective term, 
and we presented analytical solutions of illustrative boundary problems, 
which might be relevant for modeling diffusive cosmic-ray transport 
in nonuniform large-scale magnetic fields in the presence of boundaries. 
We also demonstrated the accuracy of the telegraph approximation for 
focused transport in the presence of boundaries by comparing 
the analytical solutions of the telegraph approximation with 
the numerical solutions of the original Fokker--Planck equation. 
The numerical results complement those for an infinite interval 
(Litvinenko \& Noble 2013; Effenberger \& Litvinenko 2014).

The key point in assessing the practical usefulness of the telegraph model 
is that cosmic-ray transport in turbulent magnetic fields is diffusive on
sufficiently long time scales. After a few scattering times, the
diffusion and telegraph models give very similar predictions for an
evolving particle distribution. The telegraph approximation is an
attempt to more accurately describe the particle evolution on shorter
time scales. Direct comparison of the predictions of the diffusion and
telegraph models with the numerical solution of the Fokker--Planck
equation for focused particle transport clearly shows that the
telegraph model reproduces the shape of an evolving density pulse much
better than the diffusion model, especially when focusing is strong,
even for times significantly exceeding the scattering time (see, for
instance, Figures 2 and 3 in Effenberger \& Litvinenko 2014).

In the present study, we compared the Fokker--Planck and
telegraph results for different boundary value problems and confirmed 
the validity of the telegraph approximation after just a few
scattering times. Hence the telegraph approximation can provide an improved 
description over the diffusion approximation at these
early times. Values for the parallel mean free path of about
0.1 to 0.3 AU, found in solar energetic particle studies
(e.g. Dr{\"o}ge et al. 2014), suggest that the time scale of a few
scattering times is relevant for interplanetary particle
transport, for example in the problem of predicting the
impact of large solar events at 1 AU (e.g. Shea \& Smart 2012). 

A limitation of the telegraph approximation is that solving an initial
value problem requires the knowledge of the first derivative of the
density $F_0 (z, t)$ with respect to time at $t=0$, which can be
determined only if the full distribution function $f_0 (z,\mu,v,t)$ is
known.  However, in concrete applications $\partial_t F_0 (z, 0)$ may
be known or estimated either observationally or on theoretical
grounds. For instance, in the simplest case of an isotropic initial
distribution, the derivative vanishes (see equations 4 and 5 in
Litvinenko and Schlickeiser 2014). More generally, any knowledge of
the initial distribution would yield a more accurate description of
the evolving particle distribution in the telegraph model, as compared
with the diffusion model.

Finally, the telegraph equation proved to be a useful model in a wide range 
of transport problems (for a review, see Weiss 2002; Dunkel et al. 2007), 
and so the results of the present paper should have a broad applicability.

\acknowledgments
Y.L. thanks Prof. Reinhard Schlickeiser for
supporting his visit to Ruhr-Universit\"at Bochum, where a part of
this work was completed. 
Travel support for F.E. from the International Space Science Institute
for meetings of the international team 297 is appreciated. 
R.S. acknowledges support by the Deutsche
Forschungsgemeinschaft (grant Schl 201/29-1). 


\begin{itemize}
\item{}
Artmann, S., Schlickeiser, R., Agueda, N., Krucker, S., \& Lin, R. P. 2011, 
  A\&A 535, A92

\item{}
Beeck, J., \& Wibberenz, G. 1986, 
  ApJ, 311, 437

\item{}
Bieber, J. W., Evenson, P. A., \& Matthaeus, W. H. 1987, 
  Geophys. Res. Lett., 14, 864

\item{}
Brinkman, H. C. 1956, 
  Physica, 22, 29

\item{}
Crank, J. 1984, 
  Free and Moving Boundary Problems (Oxford: Clarendon Press)

\item{}
Davies, R. W. 1954, 
  Phys. Rev., 93, 1169

\item{}
Dresing, N., G\'omez-Herrero, R., Heber, B., Klassen, A., Malandraki, O., Dr{\"o}ge, W., \& Kartavykh, Y. 2014, 
  A\&A 567, A27

\item{}
Dr{\"o}ge, W., Kartavykh, Y. Y., Klecker, B., \& Kovaltsov, G. A. 2010, 
  ApJ, 709, 912

\item{}
Dr{\"o}ge, W., Kartavykh, Y. Y., Dresing, N., Heber, B., \& Klassen, A. 2014, 
  JGR, 119, 6074

\item{}
Dunkel, J., Talkner, P., \& H\"anggi, P. 2007, 
  Phys. Rev. D, 75, 043001

\item{}
Earl, J. A. 1974, 
 ApJ, 193, 231

\item{}
Earl, J. A. 1976, 
  ApJ, 205, 900

\item{}
Earl, J. A. 1981, 
  ApJ, 251, 739

\item{}
Earl, J. A. 1992, 
  ApJ, 395, 185

\item{}
Effenberger, F., \& Litvinenko, Y. E. 2014, 
  ApJ, 783:15

\item{}
Erd\'ely, A., Magnus, W., Oberhettinger, F., \& Tricomi, F. G. 1954, 
  Tables of Integral Transforms (New York: McGraw-Hill)

\item{}
Fisk, L. A., \& Axford, W. I. 1969, 
  Sol. Phys., 7, 486

\item{}
Goldstein, S. 1951, 
  Quart. J. Mech. Appl. Math., 4, 129

\item{}
Gombosi, T. I., Jokipii, J. R., Kota, J., Lorencz, K., \& Williams, L. L. 1993, 
  ApJ, 403, 377

\item{}
Hasselmann, K., \& Wibberenz, G. 1970, 
  ApJ, 162, 1049

\item{}
He, H.-Q., \& Schlickeiser, R. 2014, 
  ApJ, 792, 85

\item{}
Hemmer, P. C. 1961, 
  Physica, 27, 79

\item{}
Hill, J. M., \& Dewynne, J. N. 1987, 
  Heat Conduction (Oxford: Blackwell Scientific)

\item{}
Jokipii, J. R. 1976, 
  ApJ, 208, 900

\item{}
Kevorkian, J. 2000, 
  Partial Differential Equations: Analytical Solution Techniques 
  (Berlin: Springer)

\item{}
Kopp, A., B\"usching, I., Strauss, R. D., \& Potgieter, M. S. 2012, 
  Comp. Phys. Comm., 183, 530

\item{}
Kunstmann, J. 1979, 
  ApJ, 229, 812

\item{}
Laitinen, T., Dalla, S., \& Marsh, M. S. 2013, 
  ApJ, 773, L29

\item{}
Lario, D., Raouafi, N. E., Kwoon, R.-Y., Zhang, J., G\'omez-Herrero, R., Dresing, N., \& Riley, P. 2014, 
  ApJ, 797, 8

\item{}
Le Roux, J. A., \& Webb, G. M. 2009, 
  ApJ, 693, 534

\item{}
Litvinenko, Y. E. 2012a, 
  ApJ, 745, 62

\item{}
Litvinenko, Y. E. 2012b, 
  ApJ, 752, 16

\item{}
Litvinenko, Y. E., \& Noble, P. L. 2013,
  ApJ, 765, 31

\item{}
Litvinenko, Y. E., \& Schlickeiser, R. 2011, 
  ApJ, 732, L31

\item{}
Litvinenko, Y. E., \& Schlickeiser, R. 2013, 
  A\&A, 554, A59

\item{}
Malkov, M., \& Sagdeev, R. 2015, 
  arXiv:1502.01799

\item{}
Marshak, R. E. 1947, 
  Phys. Rev., 71, 443

\item{}
Masoliver, J., P\`{o}rra, J. M., \& Weiss, G. H. 1992, 
  Phys. Rev. A, 45, 2222

\item{}
Masoliver, J., P\`{o}rra, J. M., \& Weiss, G. H. 1993, 
  Phys. Rev. E, 48, 939

\item{}
Pauls, H. L., \& Burger, R. A. 1994, 
  ApJ, 427, 927

\item{}
Qin, G., Wang, Y. Zhang, M., \& Dalla, S. 2013, 
  ApJ, 766, 74

\item{}
Roelof, E. C. 1969, 
  in Lectures in High-Energy Astrophysics, 
  ed. H. {\"O}gelman \& J.~R. Wayland, 111

\item{}
Sack, R. A. 1956, 
  Physica, 22, 917

\item{}
Sandroos, A., \& Vainio, R. 2007, 
  A\&A, 455, 685

\item{}
Schlickeiser, R. 2009, 
  Mod. Phys. Lett. A, 24, 1461

\item{}
Schlickeiser, R. 2011, 
  ApJ, 732, 96

\item{}
Schlickeiser, R., Artmann, S., \& Z\"oller, C. 2011, 
  Nuclear Phys. B, 212-213, 181

\item{}
Schlickeiser, R., \& Shalchi, A. 2008, 
  ApJ, 686, 292

\item{}
Schwadron, N. A., \& Gombosi, T. I. 1994, 
  J. Geophys. Res., 99, 19301

\item{}
Shalchi, A. 2011, 
  ApJ, 728, 113

\item{}
Shea, M. A., \& Smart, D. F. 2012, 
  Space Sci. Rev., 171:161

\item{}
Sofue, Y., Fujimoto, M., \& Wielebinski, R. 1986, 
  ARA\&A, 24, 459

\item{}
Wang, Y., \& Qin, G. 2015, 
  ApJ, 799, 111

\item{}
Weinberg, A. M., \& Wigner, E. P. 1958, 
  The Physical Theory of Neutron Chain Reactors 
  (Chicago: University of Chicago Press), Chap. 8

\item{}
Weiss, G. H. 2002, 
  Physica A, 311, 381

\item{}
Zhang, M., Qin, G., \& Rassoul, H. 2009, 
  ApJ, 692, 109 

\end{itemize}
\end{document}